\documentclass[a4paper,twocolumn,showpacs,superscriptaddress,amsmath,amssymb,aps,pra,preprintnumbers]{revtex4-2}
\usepackage{bbm}
\usepackage{dsfont}
\usepackage{amsmath}
\usepackage{verbatim}
\usepackage{graphicx}
\usepackage{float}
\usepackage{braket,amsmath}
\usepackage[dvipsnames, svgnames, x11names]{xcolor}
\usepackage{array}
\usepackage{bibunits}
\usepackage{braket}
\usepackage{mathrsfs}
\usepackage{amsmath}
\usepackage{graphicx}% Include figure files
\usepackage{dcolumn}% Align table columns on decimal point
\usepackage{bm}% bold math
\usepackage{hyperref}
\usepackage[capitalise]{cleveref}
\usepackage{upgreek}
\crefname{section}{Sec.}{Secs.}
\crefname{appendix}{App.}{Apps.}

\begin{document}

\emph{}
%\preprint{APS/123-QED}
\title{Enhancing collective spin squeezing via one-axis twisting echo control of individual atoms}% Force line breaks with \\
%\thanks{A footnote to the article title}

\author{Zhiwei Hu}%
\affiliation{%
State Key Laboratory of Surface Physics and Key
Laboratory of Micro and Nano Photonic Structures (Ministry of Education) and Department of Physics,
Fudan University, Shanghai 200433, China
}%
\author{Youwei Zhang}%
\affiliation{%
State Key Laboratory of Surface Physics and Key
Laboratory of Micro and Nano Photonic Structures (Ministry of Education) and Department of Physics,
Fudan University, Shanghai 200433, China
}
\author{Junlei Duan}%
\affiliation{%
State Key Laboratory of Surface Physics and Key
Laboratory of Micro and Nano Photonic Structures (Ministry of Education) and Department of Physics,
Fudan University, Shanghai 200433, China
}

%\author{Klaus Mølmer}%
%\affiliation{%
%Niels Bohr Institute, University of Copenhagen, Blegdamsvej 17, DK 2100 Copenhagen, Denmark
%}%

\author{ Mingfeng Wang}
\email{mfwang@wzu.edu.cn}
\affiliation{%
 Department of Physics, Wenzhou University, Zhejiang 325035, China
}%
\author{Yanhong Xiao}
\email{yxiao@fudan.edu.cn}
\affiliation{%
State Key Laboratory of Surface Physics and Key
Laboratory of Micro and Nano Photonic Structures (Ministry of Education) and Department of Physics,
Fudan University, Shanghai 200433, China
}%
\affiliation{%
State Key Laboratory of Quantum Optics Technologies and Devices,
Institute of Laser Spectroscopy, Shanxi University, Taiyuan, Shanxi 030006, China
}%
\affiliation{%
Collaborative Innovation Center of Extreme Optics, Shanxi University, Taiyuan, Shanxi 030006, China
}%


%%%%%%%%%%%%%%%%%%%%%%%%%%%%%%%%%%%%%%%%%%%%%%%%%%%%%%%%%%%%%%%%%%%%%%%%%%%%%%%%%%%%%%%%%%%%%%%%%%%%%%
\begin{abstract}
Spin squeezing generated via inter-atom entanglement in multilevel atomic ensembles provides a powerful resource for quantum-enhanced metrology. Existing schemes that harness internal atomic degrees of freedom to boost squeezing typically encode the collective squeezing in complex superpositions of magnetic sublevels, which complicates state control and limits practical applications. Here, we propose a coherent control scheme that simultaneously enhances collective spin squeezing and maps the resulting atom-atom entanglement onto two well-defined magnetic sublevels suitable for subsequent metrology experiments. Our protocol sandwiches a quantum non-demolition measurement between two internal one-axis-twisting interactions arranged in an echo sequence. We show that this approach can optimally leverage internal states to boost the inter-atom entanglement and, at the same time, encode it in two magnetic sublevels, which is readily convertible into metrologically useful spin squeezing. Our results offer a straightforward and efficient strategy for generating highly entangled yet readily accessible quantum states in multilevel atomic systems.
\end{abstract}
\maketitle

% Introduction
\section{Introduction}
Squeezed spin states (SSSs) of collective atomic spins can suppress quantum fluctuations below the standard quantum limit (SQL) along one quadrature \cite{PhysRevA.47.5138,physics.reports1}, providing a direct pathway to enhanced measurement precision in quantum metrology \cite{PhysRevA.46.R6797}. Fundamentally entangled on a multipartite level \cite{PhysRevA.79.042334}, atoms in SSSs form a versatile resource not only for parameter estimation surpassing the SQL---as demonstrated in atomic clocks \cite{pedrozo2020entanglement, robinson2024direct}, interferometers \cite{greve2022entanglement, malia2022distributed}, and magnetometers \cite{PhysRevLett.104.093602,PhysRevLett.113.103004,PhysRevLett.109.253605}---but also for applications in quantum information science \cite{RevModPhys.77.513,cvpp}. Therefore, the quest for highly squeezed SSSs has become a field of great interest in quantum physics.

To date, most existing studies on spin squeezing treat atoms as two-level systems, or qubits. In this framework, spin squeezing originates from inter-atom entanglement, which is commonly referred to as collective spin squeezing  \cite{PhysRevLett.104.073602,Nature60,Nature61,PhysRevLett.85.1594, appel2009mesoscopic,PhysRevLett.104.073604}. However, real atomic systems typically have multi-levels---each atom constitutes a qudit system, e.g., a spin-$f$ particle with $f>1/2$, corresponding to a qudit with dimension $d=2f+1$. Therefore, how to harness atomic internal states for spin-squeezing enhancement has emerged as a prominent and actively pursued research frontier. M\o{}lmer \emph{et al.} proposed an innovative approach to enhance spin squeezing \cite{PhysRevA.81.032314}: by first preparing atomic internal states in SSSs and then using a quantum non-demolition (QND) measurement for collective squeezing, they find that the overall degree of spin squeezing in an atomic ensemble can be increased, which has recently been experimentally demonstrated \cite{xkt1-y58b}. The key idea of this scheme is that the squeezing of internal state reduces the fluctuations of individual atomic angular momentum along a certain direction, while the collective squeezing further reduces the uncertainty of the collective angular momentum. Since both squeezing processes share the same direction, they are integrated cooperatively, resulting in an overall enhanced squeezing. However, this approach suffers a notable limitation: internal squeezing reduces the efficiency of QND detection, which in turn reduces the efficiency of collective squeezing, resulting in a total squeezing coefficient slightly lower than the direct product of the internal and collective squeezing.  Deutsch \emph{et al.} proposed a new strategy \cite{PhysRevLett.109.173603}: instead of performing the collective QND measurement along the squeezed direction of the internal spin state, they do it along the anti-squeezed direction of the individual spin. This counter-intuitive protocol can enhance the coupling strength of the QND interaction and thus improve inter-atom entanglement, which can boost the overall spin squeezing.

Despite these advancements made in the field of spin squeezing in multi-level atoms, a common challenge still remains: both approaches create ensemble squeezing by populating internal atomic states in complex superpositions. This complicates the state control required for practical applications.
For instance, in Ramsey-type spectroscopy, precise control of well-defined states---such as the preparation of coherent superpositions between two basis states---is essential \cite{RevModPhys.90.035005}.  However, for qudit atoms, the fiducial spin-up and spin-down states typically correspond to intricate superpositions of magnetic sublevels \cite{PhysRevA.81.032314,PhysRevLett.109.173603}. This complexity hinders efficient manipulation and readout of these spin states in an experiment. Therefore, developing more straightforward methods that leverage internal-state control to enhance overall spin squeezing, while directly encoding the ensemble squeezing in two-level magnetic sublevels, is of crucial importance for practical applications.

In this paper, we propose an internal-state control mechanism capable of enhancing collective squeezing while simultaneously encoding inter-atom entanglement in two magnetic sublevels. Our scheme employs a collective QND measurement inserted between two reversed internal one-axis-twisting (OAT) interactions. The first OAT evolution amplifies single-spin quantum fluctuations to strengthen the QND interaction, thereby increasing inter-atom entanglement. The second inverse OAT evolution maps this entanglement---encoded in the superposition of multiple internal states---onto two magnetic sublevels, which can then be transferred into metrologically useful spin squeezing via internal state exchange. We show that such a twisting echo protocol enables the full utilization of the internal degrees of freedom for squeezing enhancement. We also analyze the performance of the scheme in the presence of decoherence, showing that the echo-control protocol effectively enhances the optical depth (OD) of an atomic ensemble by up to a factor of $2f$. Therefore, our echo protocol offers a promising solution for achieving a high degree of spin squeezing in atomic systems with a small OD but large $f$.

The rest of the paper is organized as follows. In \cref{sec2}, we introduce the physical system and the basic principle of our scheme. In \cref{sec3}, we quantify the squeezing produced by the proposed scheme. In \cref{sec4}, we present the implementation of the scheme and analyze the effects of imperfection. Finally, a summary is provided in \cref{sec5}.

\section{The model\label{sec2}}
\begin{figure}[t]
	\centering
	\includegraphics[width=1\linewidth]{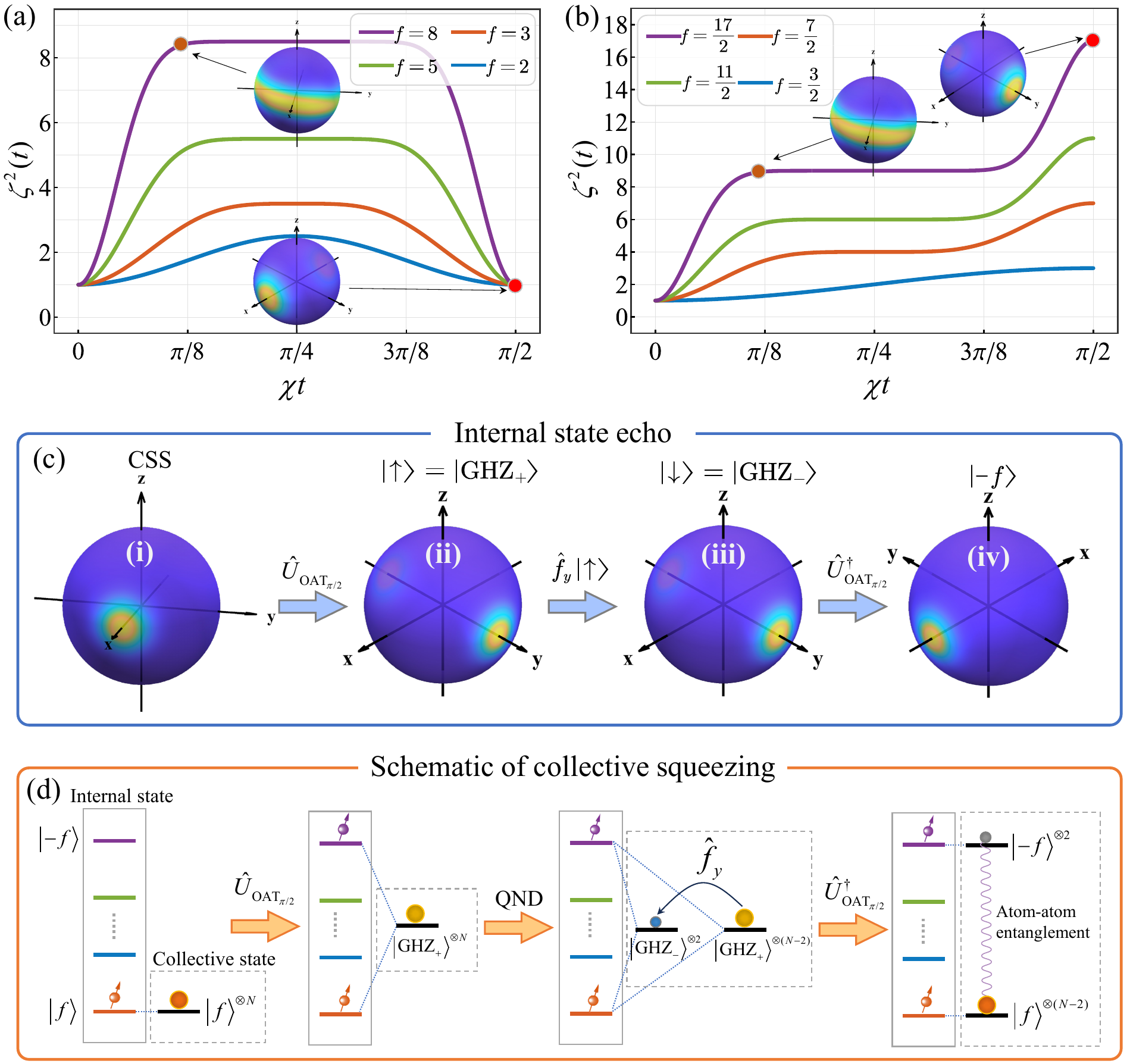}
	\caption{Dependence of the enhancement factor on the coupling strength for (a) integer $f$ and (b) half-integer $f$.
(c) Schematic of the internal-state twisting echo protocol for enhancing inter-atom entanglement, depicted on a Bloch sphere. Initial CSS (i) evolves under $\hat U_{\rm{OAT}}$ into a GHZ state $\ket{\rm{GHZ_+}}$ (ii). The QND interaction drives a transition from $\ket{\rm{GHZ_+}}$ to $\ket{\rm{GHZ_-}}$ (iii), while both states exhibit the same probability distribution on the Bloch sphere. Finally, the reversed OAT evolution $\hat U_{\rm{OAT}}^\dag$ transforms $\ket{\rm{GHZ_-}}$ ($\ket{\rm{GHZ_+}}$) back into the magnetic sublevel $\ket{-f}$ ($\ket{f}$) (iv). (d) Schematic of collective squeezing. The first OAT evolution transforms the collective state from the reference state $\ket{f}^{\otimes N}$ to $\ket{\rm{GHZ}_+}^{\otimes N}$. The subsequent QND measurement then coherently transfers an even number of atoms to $\ket{\rm{GHZ}_-}$, generating inter-atom entanglement encoded in the two orthogonal $\rm{GHZ}_\pm$ states. This entanglement is finally mapped onto the magnetic sublevels $\ket{-f}$ and $\ket{f}$ via inverse OAT evolution.}
	\label{fig1}
\end{figure}

We consider an atomic ensemble of $N$ identical spin-$f$ atoms. Each atom is initialized in the magnetic sublevel $\ket{f}$ (quantization axis $x$), where $\ket{m_f^x}\equiv\ket{f,m_f^x}$ denotes an eigenstate of single-atom $x$-component $\hat f_x$. Collectively, the atoms form a coherent spin state (CSS) $\ket{\rm{CSS}}=\ket{f}^{\otimes N}$ [\cref{fig1}(c)(i)]. The core concept behind the enhancement of collective squeezing via qudit control given in \cite{PhysRevLett.109.173603} is to maximize the uncertainty of internal state. To do so, we employ the single-atom OAT interaction
\begin{eqnarray}
 \hat H_{\rm{OAT}}=\chi \hat f_z^2,
 \end{eqnarray}
 where $\chi$ denotes the coupling constant. The dynamics generated by $\hat H_{\rm{OAT}}$ gradually stretches the uncertainty associated with $\hat f_y$ \cite{PhysRevA.47.5138}. At the specific time $t=\pi/2\chi$ \cite{PhysRevLett.82.1835},
 the dynamics produce a maximally entangled GHZ state (effective entanglement of electronic spin and nuclear spin \cite{PhysRevLett.101.073601}), $\ket{f,m^y_f=f}+e^{i\pi f}\ket{f,m^y_f=-f}$, which has Heisenberg-limited (HL) maximal uncertainty \cite{PhysRevA.56.2249,evrard2019enhanced,jk8g-t1d8}. We note that this is an internal GHZ state of a single atom, also referred to as cat state or kitten state \cite{NPNPX}.

  The proposed twisting echo protocol is shown in \cref{fig1}(c). First, we apply an OAT interaction to an individual atom for time $t$, obtaining the reference state  $\ket{\uparrow}\equiv\hat U_{\rm{OAT}}\ket{\rm{CSS}}=\sum_{k=0}^fc_k\ket{f-2k}$, where $\hat U_{\rm{OAT}}=e^{-i\chi t\hat f_z^2}$ and $c_k=\langle{f-2k}|{\uparrow}\rangle$. This twisted state contains only an even number of spin excitations and corresponds to a SSS at short interaction times $t$, with a squeezing direction that varies over time \cite{PhysRevA.47.5138}. Next, we measure the anti-squeezed quadrature (e.g., $y$) rather than the squeezed quadrature \cite{xkt1-y58b}, employing the QND interaction described by
\begin{eqnarray}
      \hat{H}_{\rm{QND}}^y=\alpha\hat F_y\hat S_z,
\end{eqnarray}
where $\alpha$ is the coupling constant, ${\hat F_{x,y,z}} = \sum\nolimits_{i = 1}^N {\hat f_{x,y,z}^{(i)}}$ are the collective angular momentum operators, and $\hat S_{y}=\frac{1}{2i}\int_0^Tdt(\hat a_+^\dag \hat a_--\hat a_-^\dag \hat a_+)$, $\hat S_{z}=\frac{1}{2}\int_0^Tdt(\hat a_+^\dag \hat a_+-\hat a_-^\dag \hat a_-)$ are the collective Stokes operators for a light pulse of duration $T$, with $\hat a_\pm$ denoting the annihilation operators for the $\sigma_\pm$ polarization modes \cite{PhysRevLett.85.5643}. For a strong $z$-polarized light pulse, one may apply the Holstein-Primakoff approximation to Stokes operators by defining the light quadratures $( {{{\hat X}_L},{{\hat P}_L}} ) =( {{{\hat S}_y},{{\hat S}_z}})/\sqrt {{N_p}/2}$ with $N_p$ being the photon number of the pulse, obeying $[ {{{\hat X}_L},{{\hat P}_L}} ] = i$ \cite{PhysRevLett.97.143602}. Assuming that the $y$-polarized optical mode is in the vacuum state $\ket{0}_L$, the quantum state of the system under weak coupling is given by
 \begin{eqnarray}
{{\hat U}_{{\rm{QND}}}}{{\left| 0 \right\rangle _L}\left|  \uparrow  \right\rangle ^{ \otimes N}} &\approx&  \left[ {\left( {1 - \frac{1}{4}{{\tilde \alpha }^2}\hat F_y^2} \right){{\left| 0 \right\rangle }_L} + \frac{{\tilde \alpha }}{{\sqrt 2 }}{{\hat F}_y}{{\left| 1 \right\rangle }_L}} \right.\nonumber\\
&&\left. { + \frac{{{{\tilde \alpha }^2}}}{{2\sqrt 2 }}\hat F_y^2{{\left| 2 \right\rangle }_L}} \right]{\left|  \uparrow  \right\rangle ^{ \otimes N}}, \label{eq1}
\end{eqnarray}
where ${{\hat U}_{{\rm{QND}}}} = {e^{ - i\tilde \alpha {{\hat F}_y}{{\hat P}_L}}}\approx{1 - i\tilde \alpha {{\hat F}_y}{{\hat P}_L} - {{\tilde \alpha }^2}\hat F_y^2\hat P_L^2}/2$ for small $\tilde \alpha  = \sqrt {{N_p}/2} \alpha$ and $\ket{n}_L$ represents the Fock state of the optical mode. Next, a homodyne detection of the optical position quadrature is performed. If the measurement yields the result $x_m$, the quantum state collapses to: $\langle {x_m}|{\hat U_{{\rm{QND}}}}{\left| 0 \right\rangle _L}{\left|  \uparrow  \right\rangle ^{ \otimes N}} \approx [1 + \tilde \alpha {x_m}{\hat F_y} + {\tilde \alpha ^2}\left( {x_m^2 - 1} \right)\hat F_y^2/2]{\left|  \uparrow  \right\rangle ^{ \otimes N}}$, where we have used the relation $\bra{x_m}n\rangle_L  = {e^{ - x_m^2/2}}{H_n}({x_m})/\sqrt {{2^n}n!\sqrt \pi  }$, with $H_n(.)$ being the Hermite polynomial, and have omitted the normalization coefficient. For a specific measurement result $x_m = 0$, yields the atomic state  \cite{PhysRevLett.109.173603}
\begin{eqnarray}
{\left| \Psi  \right\rangle _A} \approx | \uparrow {\rangle ^{ \otimes N}} - \frac{{{\kappa ^2}{{\zeta}^2}}}{{4N}}\sum\limits_{{\rm{perm }}} |  \downarrow {\rangle ^{ \otimes 2}}| \uparrow {\rangle ^{ \otimes (N - 2)}},\label{eq2}
\end{eqnarray}
where we have defined the coupling strength $\kappa=\sqrt{fN}\tilde{\alpha}$, the enhancement factor $\zeta={\Delta {f_y}\sqrt{2/f}}$, and the coupled state $\ket{\downarrow} \equiv {{\hat f}_y}\ket{\uparrow}/\Delta {f_y} = \sum\nolimits_{k = 0}^{f - 1} {{d_k}\ket{{ f - (2k+1)} } }$ with $\Delta f_y$ being the standard
deviation of $\hat f_y$ in the fiducial state $\ket{\uparrow}$ and $d_k=\langle{f-(2k+1)}|{\downarrow}\rangle$ \cite{xkt1-y58b}. It is the pairwise entanglement in Eq. (\ref{eq2}) that gives rise to the collective spin squeezing. Normally, the stronger the inter-atom entanglement, the higher the degree of collective spin squeezing. Notably, the amount of inter-atom entanglement depends on $\zeta$---a parameter proportional to the variance of $\hat f_y$ in the reference state $\ket{\uparrow}$. Therefore, increasing the variance of $\hat f_y$ is a straightforward strategy for enhancing inter-atom entanglement and, ultimately, boosting spin squeezing \cite{PhysRevLett.109.173603}. From the reference state $\ket{\uparrow}$, one may derive the expectation values at time $t$ \cite{PhysRevA.47.5138}
\begin{eqnarray}
\langle {{{\hat f}_y}}\rangle  = 0,\langle {\hat f_y^2}\rangle  = \frac{f}{2}\left[ {f + \frac{1}{2} - \left( {f - \frac{1}{2}} \right){{\cos }^{2f - 2}}(2\chi t)} \right],\nonumber
\end{eqnarray}
and therefore the enhancement factor
\begin{eqnarray}
\zeta^2 \left( t \right) = f + \frac{1}{2} - \left( {f - \frac{1}{2}} \right){\cos ^{2f - 2}}(2\chi t).\label{eq3}
\end{eqnarray}
In a standard QND scheme, the reference state is a CSS, yielding $\zeta^2(0)=1$. Consequently, the efficiency of collective squeezing is determined solely by the QND coupling strength $\kappa$ \cite{Nature.581.7807}. However, if the internal state is anti-squeezed along the $y$ direction, as shown in the insets of \cref{fig1}(a) and (b), then $\zeta> 1$. As a result, the coupling strength is effectively increased to $\kappa\zeta$, thereby enhancing the collective spin squeezing. It is worth noting that the value of $\zeta$ depends on whether $f$ is an integer or a half-integer. For large integer $f$, $\zeta^2$ rapidly reaches its maximum value of $f + 1/2$ at time around $t\approx 1/(\sqrt{f}\chi)$ and exhibits a long plateau thereafter. When $\chi t>\pi/4+1/\sqrt{f}$, $\zeta^2$ plummets back to $1$. For half-integer $f$, however, $\zeta^2$ grows further after entering the plateau and reaches a maximum of $2f$ at $\chi t = \pi/2$, which corresponds to a GHZ state, shown in the inset of Fig. 1(b).

Although the reference states $\ket{\uparrow}$ within the plateau region [see Fig. \ref{fig1}(a) and (b)] enable enhanced collective squeezing, the generated squeezing (or atom-atom entanglement) is encoded in complex internal superpositions, which hinders practical applications. To transfer the superposition state $\ket{\uparrow}$ back into one magnetic sub-state, we propose to apply a reverse OAT evolution, $\hat U_{\rm{OAT}}^\dag$, to individual atoms, resulting in
\begin{eqnarray}
{\left| \Psi  \right\rangle _{A1}} \approx |f{\rangle ^{ \otimes N}} - \frac{{{\kappa ^2}}}{{2Nf}}\sum\limits_{{\rm{perm}}} {{\ket{\downarrow'}^{\otimes 2}}|f{\rangle ^{ \otimes (N - 2)}}},\label{eq4}
\end{eqnarray}
where the spin-down state (see Appendix A for more details) is given by
 \begin{eqnarray}
\ket{\downarrow'} &=& \hat U_{\mathrm{OAT}}^\dag \hat f_y \hat U_{\mathrm{OAT}} \ket{f}\nonumber\\
    &=& e^{-i\chi t} \left[f \sin \bigl( 2\chi t \hat f_z \bigr) \ket{f}
    + i\sqrt{\frac{f}{2}} \cos \bigl( 2\chi t \hat f_z \bigr) \ket{f-1}\right].\label{replyeq7}\nonumber\\
\end{eqnarray}
For half-integer $f$, at the particular time $t=\pi/2\chi$, the coupled state evolves into the magnetic sub-state $\ket{\downarrow'}=fe^{-i\pi f}\ket{-f}$ (see Appendix A). Consequently, the atomic state of Eq. (\ref{eq4}) becomes
\begin{eqnarray}
{\left| \Psi  \right\rangle _{A2}} \approx |f{\rangle ^{ \otimes N}} + \frac{{{(\sqrt{2f}\kappa) ^2}}}{{4N}}\sum\limits_{{\rm{perm}}} {{\ket{-f}^{\otimes 2}}|f{\rangle ^{ \otimes (N - 2)}}}.\label{eq5}
\end{eqnarray}
This is the main result of this work. It demonstrates that the QND coupling strength $\kappa$ is enhanced by a factor of $\sqrt{2f}$, thereby promoting atom-atom entanglement. As expected, this entanglement is naturally encoded between the magnetic sub-states $\ket{f}$ and $\ket{-f}$. We refer to such an enhancement as HL enhancement because the internal GHZ state is the state with maximum spin uncertainty.
In \cref{fig1}(c) and (d), we schematically illustrate the procedure for generating enhanced collective squeezing. The initial CSS is first evolved into the reference state, $\left| {{\rm{GH}}{{\rm{Z}}_ + }} \right\rangle  = (\ket{{f,m_f^y = f}} + e^{i\pi f}\ket{{f,m_f^y =  - f}} )/\sqrt 2$, via the OAT dynamics (see Appendix B) \cite{PhysRevA.56.2249}. Subsequently, a QND measurement couples the reference state to $\left| {{\rm{GH}}{{\rm{Z}}_ - }} \right\rangle  = (\ket{{f,m_f^y = f}} - e^{i\pi f}\ket{{f,m_f^y =  - f}} )/\sqrt 2$. Then, a reverse OAT evolution is applied, transforming atoms in the $|\rm{GHZ}_+\rangle$ state back into the $|f\rangle$ state, while those in the $|\rm{GHZ}_-\rangle$ state are transformed into the $|-f\rangle$ state. This process finally yields the entangled state in Eq. (\ref{eq5}).

We would like to stress that our proposed scheme is also applicable to integer values of
 $f$. In this scenario, rather than detecting $\hat F_y$, one should measure $\hat F_x$ through the QND interaction $\hat H_{\rm{QND}}^x=\alpha\hat F_x\hat S_z$. This is because the GHZ state created by OAT dynamic is oriented along the $x$-axis (which is the direction of maximum spin uncertainty), as is evident from the inset of \cref{fig1}(a). Although the required measurement axis differs, the internal-state manipulation procedure itself is identical to that for half-integer $f$, and the resulting entangled atomic state is the same as that given in Eq. (\ref{eq5}).

 \section{Spin squeezing\label{sec3}}
 To quantify the enhanced collective squeezing, we utilize a multilevel Holstein-Primakoff approximation \cite{PhysRevA.81.032314,xkt1-y58b}. For the reference state $\ket{\uparrow}$, one can assign to $\hat F_y$ an oscillator quadrature variable \cite{PhysRevA.81.032314}: $\hat X_A = \hat F_y / \sqrt{ 2 ( \Delta \hat F_y )_{\ket{\uparrow}}^2 }$, where $( {\Delta \hat O} )_{| {\uparrow} \rangle }^2$ denotes the variance of $\hat O$ associated with the reference state $\ket{\uparrow}$. As a result, the QND interaction can be rewritten as $\hat H_{\rm{QND}}^y=\zeta\kappa\hat X_A\hat P_L$. Under this interaction,  evaluating the input-output relations for light and atoms in the Heisenberg picture yields
\begin{eqnarray}
\hat X_L^{{\rm{out}}} = \hat X_L^{{\rm{in}}} + \zeta \kappa \hat X_A^{{\rm{in}}},\hat X_A^{{\rm{out}}} = \hat X_A^{{\rm{in}}},
\end{eqnarray}
where the superscript `in' (`out') denotes the variables before (after) the interaction. Homodyne detection of $\hat X_L^{\rm out}$ provides information about $\hat X_A$, thus reducing its variance. To verify this reduction, a feedback control displaces $\hat X_A$ by an amount proportional to the measurement result, yielding the outcome \cite{PhysRevA.72.052313}: $
 {( {\Delta \hat X_A^{{\rm{out}}}} )^2} = {g^2}/2 + {( {1 + g\zeta \kappa } )^2}/2$,
  where $g$ is the feedback gain, and we have used the result $(\Delta\hat X_A^{\rm{in}})^2=1/2$ for the initial vacuum state (that is, the reference state $\ket{\uparrow}^{\otimes N}$ \cite{PhysRevA.81.032314}). The final spin variance can be minimized to yield the squeezing parameter for the atomic oscillator
  \begin{figure}[t]
	\centering
	\includegraphics[width=1\linewidth]{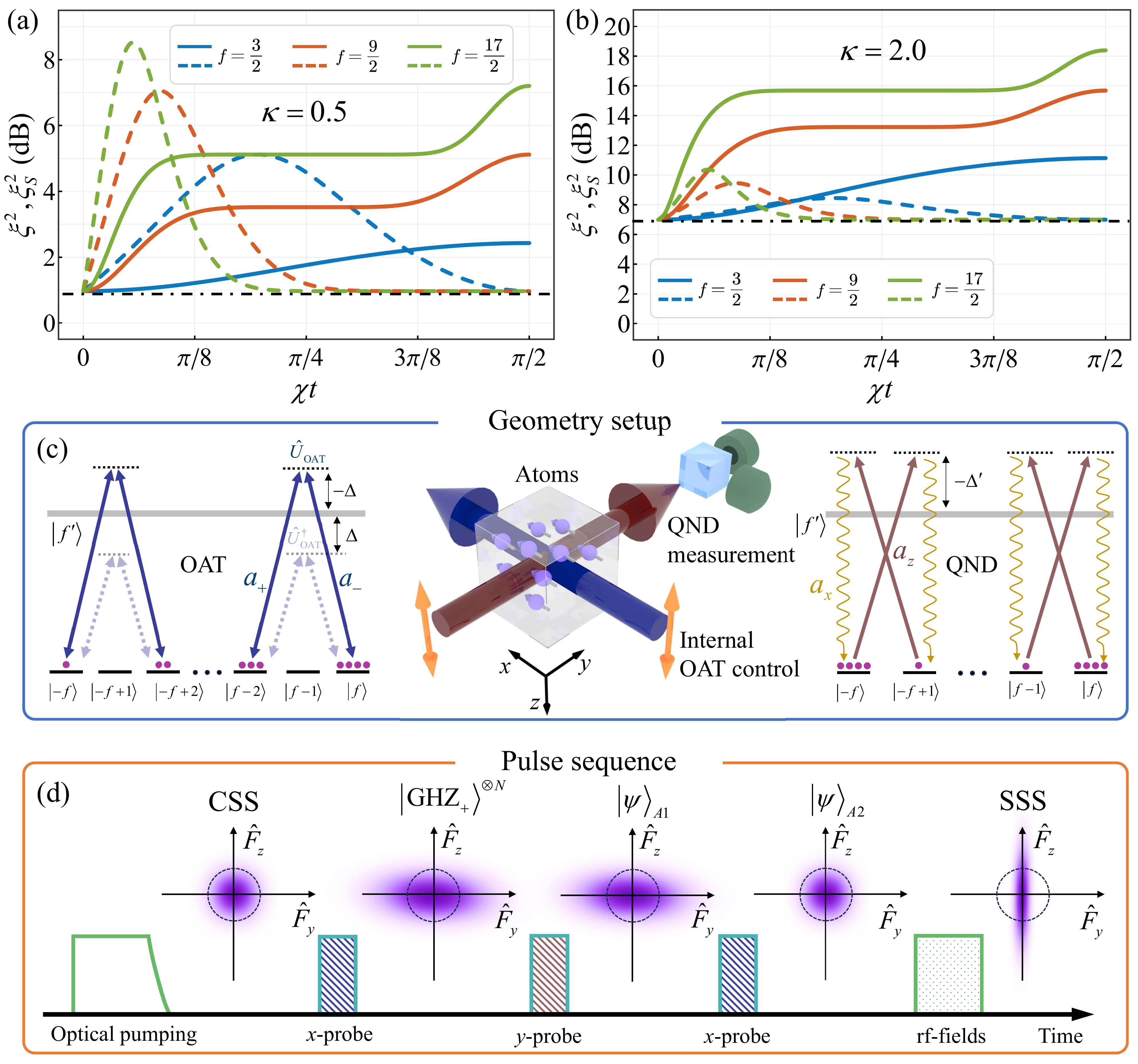}
	\caption{Spin squeezing versus internal OAT coupling strength for (a) weak and (b) strong QND measurement. Dashed curves: the cooperative squeezing scheme; solid curves: the present scheme; dot-dashed lines: the spin squeezing achieved for the internal state in CSS. (c) Schematics for implementing internal OAT and collective QND interactions (middle). The linearly polarized probe pulse propagating along the $x$-axis drives a $\Delta m_f=2$ transition (left), realizing the internal $\hat U_{\rm{OAT}}$ evolution (blue detuning) and  the $\hat U_{\rm{OAT}}^{\dag}$ evolution (red detuning). The pulse propagating along the $y$-axis is used to drive the $\Delta m_f=1$ transition (right) and responsible for the collective QND measurement. (d) Pulse sequence. Atoms are first prepared in the CSS by optical pumping, followed by the first $x$-probe, which spreads the quantum uncertainty along the $y$-direction. The subsequent $y$-probe slightly squeezes $\hat F_y$. The second $x$-probe untwists the spin to restore the variance of $\hat F_y$ to the CSS level. Finally, the rf-fields convert the atomic oscillator squeezing into spin squeezing, thereby reducing the uncertainty in $\hat F_y$. }
	\label{fig2}
\end{figure}
\begin{eqnarray}
{\xi ^2} = 2\left( {\Delta \hat X_A^{{\rm{out}}}} \right)_{\min }^2 = \frac{1}{{1 + {\zeta^2}{\kappa ^2}}},\label{eq8}
\end{eqnarray}
for $g =  -\zeta \kappa /(1 + {\zeta ^2}{\kappa ^2})$. This result differs significantly from that obtained via cooperative internal and collective squeezing \cite{xkt1-y58b}, where a QND measurement is performed on the squeezed quadrature. Although measuring along the squeezing direction of the internal state can enhance the overall degree of squeezing, such measurement, however, reduces the efficiency of QND measurement, resulting in an overall spin squeezing given by
\begin{eqnarray}
 \xi _S^2 = \frac{\xi _{{\rm{OAT}}}^2}{1 + \xi _{{\rm{OAT}}}^2{\kappa ^2}},
  \end{eqnarray}
  where $\xi _{{\rm{OAT}}}^2$ denotes the internal squeezing created by OAT evolution \cite{xkt1-y58b}. Obviously, it is worse than the product of the squeezing coefficients of internal and collective squeezing. Specifically, in strong-coupling regime ($\kappa\gg 1$), the internal squeezing offers no advantage, as $\xi_S^2\approx 1/\kappa^2$, which is the spin squeezing achievable by a conventional QND scheme with the internal state in a CSS \cite{PhysRevA.70.044304}. In contrast, for the result in Eq. (\ref{eq8}), one approximately has $\xi^2\approx 1/(f\kappa^2)$ at $t\approx 1/(\sqrt{f}\chi)$, which is equivalent to squeezing the internal state near the HL and then combining it with external-state squeezing, effectively yielding a direct product of the two types of squeezing.

Figures \ref{fig2}(a) and \ref{fig2}(b) show the squeezing produced by the two schemes as a function of time for different coupling strengths. One can clearly see that the cooperative squeezing scheme exhibits superior performance in the weak coupling regime, whereas the present scheme is more suitable for large $\kappa$. This is because the former scheme squeezes both the internal and collective states simultaneously; even when the coupling strength
$\kappa$ is small and the collective contribution is limited, the internal squeezing remains robust, ensuring a favorable overall squeezing. In contrast, the latter scheme only squeezes the collective state, so its efficiency is directly constrained by the strength of $\kappa$. In fact, provided the total internal spin $f$ is sufficiently large, the squeezing performance of the present scheme can surpass that of the cooperative scheme even in the weak coupling regime, as the scaling of the squeezing parameter $\xi^2\approx1/(f\kappa^2)\propto f^{-1}$  outperforms $\xi_S^2\approx\xi^2_{\rm{OAT}}\propto f^{-2/3}$ \cite{PhysRevA.47.5138}. The physical mechanism underlying the enhancement of inter-atom entanglement via the measurement of the spin component with large fluctuations can be understood as follows: the quantum states exhibiting large fluctuations are more susceptible to transitions from the reference state $\ket{\uparrow}$ to the coupled state $\ket{\downarrow}$ under the QND interaction.

It should be stressed that the squeezing of the atomic oscillator quadrature $\hat X_A$ is not equivalent to spin squeezing. For a SSS, the Wineland criterion \cite{PhysRevA.50.67}, defined as $\xi_W^2 =2 f N(\Delta \hat F_{\perp})^2 /\langle\hat{F}_x\rangle^2$, quantifies the amount of metrologically useful entanglement (i.e., relevant for quantum-enhanced magnetometry), where $(\Delta \hat{F}_{\perp})^2$ denotes the minimum variance in a direction perpendicular to the direction of the mean of the collective spin $\hat F_x$. We note that, for the maximum achievable squeezing of $\xi^2_{\rm{max}}=1/(1+2f\kappa^2)$, which is produced by initially preparing individual atoms in the reference state $\ket{\rm{GHZ_+}}^{\otimes N}$, the macroscopic spin vanishes as $\langle \hat F_x\rangle=0$. However, the atomic polarization can be recovered by an inverse OAT evolution as described above, yielding the atomic state of Eq. (\ref{eq5}). Although the state in Eq. (\ref{eq5}) possesses a macroscopic spin (since most atoms remain in $\ket{f}$, only a small fraction is excited to $\ket{-f}$ by the QND measurement), it is still not an SSS, since $(\Delta \hat F_{\perp})^2=Nf/2$ and therefore $\xi_W^2 =1$ [see \cref{fig2}(d)]. Nevertheless, the squeezing of the atomic oscillator can be converted into the spin squeezing by applying a sequence of $\sigma^+$-polarized rf $\pi$ pulses (or Raman transitions \cite{PRXQuantum.6.020352}) that transfer atoms from $\ket{-f}$ to $\ket{f-1}$ while leaving the atoms in $\ket{f}$ unchanged, resulting in $\xi^2_W=\xi^2_{\rm{max}}$.

\section{Implementation and imperfection\label{sec4}}
\subsection{Experimental implementation}
To generate internal OAT evolution, we consider a dispersive interaction between atoms and light propagating in the $x$ direction, as illustrated in \cref{fig2}(c). Under off-resonant conditions, the population of the excited levels is small, allowing the excited states to be adiabatically eliminated, yielding the effective interaction Hamiltonian for the $i$th atom \cite{PhysRevA.81.032314,RevModPhys.82.1041}
\begin{eqnarray}
\hat H_{AL}^{(i)} &=& \frac{g}{\Delta }\left\{ {{a_0(\Delta)}\hat \phi (0) + {a_1(\Delta)}{{\hat s}_z}(0)\hat f_x^{(i)}} \right.\nonumber\\
 &&+ {a_2(\Delta)}\left[ {-\hat \phi (0)\hat f_x^{(i)2} + 2{{\hat s}_x}(0)\left( {\hat f_y^{(i)2} - \hat f_z^{(i)2}} \right)} \right.\nonumber\\
&&\left. {\left. { + 2{{\hat s}_y}(0)\left( {\hat f_z^{(i)}\hat f_y^{(i)} + \hat f_y^{(i)}\hat f_z^{(i)}} \right)} \right]} \right\},\label{eq9}
\end{eqnarray}
where $g$ is the effective coupling constant and $\Delta$ is the detuning between ground and excited state. Here, $a_0$, $a_1$, and $a_2$ are the coefficients for the scalar, vector, and tensor polarizabilities, respectively, each of which depends on the detuning $\Delta$. $\hat \phi$ and $\hat s_{x,y,z}$ denote the spatially localized photon flux and Stokes operators. For linear polarization along the $z$-direction, the Stokes $x$ component has a macroscopic expectation value, $\hat s_x=-\phi/2$, while the other two Stokes operators, $\hat s_y$ and $\hat s_z$, are negligible. Consequently, Eq. (\ref{eq9}) simplifies to
\begin{eqnarray}
\hat H_{{\rm{OAT}}}^{(i)} \simeq 2g\phi\frac{{ {a_2(\Delta)}}}{\Delta }\hat f_z^{(i)2}.
\end{eqnarray}
Therefore, one may realize internal-state OAT evolution by simply sending a linearly polarized, off-resonant optical pulse through the atomic ensemble along the macroscopic spin, and the inverse OAT evolution can be achieved by switching the sign of the detuning ($\Delta\rightarrow-\Delta$)---readily accomplished by changing from blue to red detuning [see \cref{fig2}(c)]. For example, for \(^{87}\mathrm{Rb}\) atoms, by choosing the red and blue detunings to have the same magnitude, \(\Delta=\pm 2\pi\times 479.5~\mathrm{MHz}\), one obtains OAT and inverse-OAT interaction strengths that are equal in magnitude and opposite in sign (see Appendix C).
 A QND interaction can be realized by sending a $z$-polarized optical pulse along the $y$-direction. As the propagation direction of the probe light is changed, the spin components in Hamiltonian (\ref{eq9}) transform as \(\hat f_x \to \hat f_y\) and \(\hat f_y \to -\hat f_x\).
Under far-off-resonance conditions (with detuning $\Delta'$ ), the tensor part in Eq. (\ref{eq9}) becomes negligible (e.g. $a_1(\Delta') \rightarrow 1/24$, and $a_2(\Delta') \rightarrow 0$ for $^{133}$Cs \cite{RevModPhys.82.1041}) and thus Eq. (\ref{eq9}) reduces to
$\hat H_{{\rm{QND}}}^{(i)} \simeq \frac{{g{a_1}}}{\Delta'}{{\hat s}_z}(0)\hat f_y^{(i)}$, which yields $\hat H_{QND}^y = \sum\nolimits_{i=1}^N {\hat H_{{\rm{QND}}}^{(i)}}$.

Figure \ref{fig2}(d) illustrates the pulse sequence used to implement the proposed scheme. After preparing a CSS via optical pumping, an $x$-probe is sent through the atoms to increase $\hat F_y$ variance. This is followed by a $y$-probe, which performs a QND measurement of $\hat F_y$. However, this measurement does not significantly reduce the variance of $\hat F_y$, because both the reference state  $\ket{\uparrow}$ and the coupled state $\ket{\downarrow}$ have large uncertainties along the
$y$-direction. Subsequently, another $x$-probe is applied to realize an inverse OAT evolution, transforming both the reference and coupled states back into CSSs. Finally, the radio-frequency fields map $\ket{-f}\rightarrow\ket{f-1}$, and subsequent feedback control (also implemented via an additional rf pulse \cite{PhysRevLett.110.163602}) yields an enhanced SSS. We stress that the feedback control can be applied either immediately after the QND measurement or at the end of the protocol. The latter is experimentally more convenient, as it only requires a rotation of the collective spin.

\subsection{Noise effect}

The performance of the proposed protocol is inevitably influenced by decoherence. Given that high-fidelity control of the internal state is achievable \cite{PRXQuantum.6.020352,PhysRevA.104.L060401}, we here concentrate on the noise effect arising from collective squeezing. The fundamental source of collective decoherence is spontaneously emitted light scattered into other forward scattering modes that will not be detected by photodetectors \cite{RevModPhys.82.1041}. This process leads to the decay of collective spin at a rate $\beta/2T$, modifying the evolution of atoms by introducing a decay term \cite{hammerer2006quantum,PhysRevA.72.052313,PhysRevLett.85.5643}: ${\partial_t}{{\hat X}_A} =  - {\beta}/({2T}){{\hat X}_A} + \sqrt {{\beta}/{T}} {{\hat f}_{XA}}$,
 where $\hat f_{XA}$ is the standard vacuum noise operator with zero mean and correlation $\langle\hat f_{XA}(t)\hat f_{XA}(t')\rangle=\delta(t-t')/2$. Regarding the probe light, dissipation of the light quadrature arises from both atomic absorption and light reflections by photon detectors, modeled as \cite{PhysRevA.74.064301,PhysRevA.70.044304}: $
 \hat X_L^{{\rm{out'}}} = \sqrt {1 - \epsilon } [\hat X_L^{{\rm{in}}} + \zeta \kappa \int_0^T {dt{{\hat X}_A}(t)/T } ] + \sqrt \epsilon  {{\hat F}_{LX}}$,
  where $\epsilon$ denotes the damping coefficient and ${{\hat F}_{LX}}$ is the optical vacuum operator satisfying $\langle{{\hat F}_{LX}}^2\rangle=1/2$. In the presence of these noises, the squeezing parameter becomes (see Appendix D for more details)
   \begin{eqnarray}
  {\xi ^2} = \frac{1}{1 + (1 -\epsilon ){\zeta ^2}{\kappa ^2}} + \frac{\beta}{3},\label{Feq14}
    \end{eqnarray}
   indicating that light loss effectively reduces the coupling strength by a factor $\sqrt{1-\epsilon}$, while the atomic decay sets an upper bound on the maximum achievable squeezing. In fact, the coupling strength can also be expressed as $\kappa^2=\alpha_0\beta$ \cite{RevModPhys.82.1041}, where $\alpha_0$ is the OD of the atomic ensemble. Consequently, the squeezing parameter reaches a minimum of
    \begin{eqnarray}
   {\xi ^2_{\rm{min}}}\approx \frac{2}{\sqrt{3(1-\epsilon)\zeta ^2\alpha_0}},\label{eq18}
    \end{eqnarray}
    obtained at $\beta=\sqrt{3/(1-\epsilon)\zeta ^2\alpha_0}$.
  Eq. (\ref{eq18}) demonstrates that the internal OAT control effectively enhances the OD by a factor of $(1-\epsilon)\zeta ^2$.  This scaling indicates that the advantage of our scheme becomes increasingly significant in high-spin platforms. For example,  ${}^{167}\mathrm{Er}$ possesses hyperfine manifolds up to $f=19/2$ \cite{PhysRevA.88.032508}, for which substantially enhanced spin squeezing may be achievable.

\section{Conclusion\label{sec5}}
We proposed a scheme for enhanced collective spin squeezing based on a QND measurement sandwiched between two internal OAT evolutions. Unlike conventional approaches that rely on the squeezing dynamics of OAT, our scheme actively uses its anti-squeezing property: the first OAT evolution deliberately amplifies the uncertainty of individual spin, and a subsequent QND measurement on the amplified spin component then significantly enhances the effective atom-light interaction strength, thereby promoting inter-atom entanglement and amplifying the squeezing of the atomic oscillator. To convert this oscillator squeezing into spin squeezing, we apply a reverse OAT evolution. We demonstrate that if the initial OAT evolution prepares the atoms in a GHZ state, the subsequent reverse evolution perfectly maps the entanglement encoded in the superposition of internal states onto the two magnetic sublevels. Finally, a straightforward state exchange between these sublevels yields the enhanced spin squeezing.

Our scheme fully exploits the internal-state degree of freedom for spin-squeezing enhancement, and the resulting SSSs are directly applicable in quantum metrology \cite{RevModPhys.90.035005}, quantum memory \cite{PhysRevLett.101.073601} and quantum information science \cite{RevModPhys.77.513}. When accounting for noise effects, we find that internal-state control effectively enhances the OD of an atomic ensemble by up to a factor of $2f$. Such enhancement suggests potential of superior performance for more efficient squeezing mechanisms. For instance, in the case of two-axis twisting squeezing dynamics \cite{PhysRevA.96.013823,luo2025hamiltonian,miller2024two}, the scaling of the squeezing parameter can be improved from $\propto1/\alpha_0$ to $\propto1/(2f\alpha_0)$, corresponding to an enhancement by a HL amount of  $1/(2f)$. A direct extension of this work is to measure a static magnetic field via Ramsey interferometry \cite{NPNPX} or to transfer squeezing onto clock transitions \cite{pedrozo2020entanglement} using tailored composite pulses \cite{25ds-9724,NNNN223}.

\begin{acknowledgments}

We acknowledge Zhenguo Wang for preparing the schematic illustrations. This work is supported by STCSM24LZ1400400, the Innovation Program for Quantum Science and Technology under Grant No.~2023ZD0300900, the Natural Science Foundation of China (NSFC) under Grants No.~12161141018, No.~12027806, Fund for Shanxi ``1331 Project", and the Zhejiang Provincial Natural
Science Foundation of China (Grant No. LMS25A040004).
\end{acknowledgments}
\bibliography{main}% Produces the bibliography via BibTeX.
\appendix
\section{Derivation of the spin-down state\label{AppA}}
In this Appendix, we present a detailed derivation of the spin-down state $\ket{\downarrow'} = \hat U_{\mathrm{OAT}}^\dag \hat f_y \hat U_{\mathrm{OAT}} \ket{f}$ introduced in the main text.
For the OAT evolution operator
\(\hat U_{\mathrm{OAT}}=e^{-i\chi t \hat f_z^2}\),
the single-spin ladder operators \(\hat f_\pm=\hat f_x \pm i\hat f_y\) transform as \cite{PhysRevA.47.5138}
\begin{align}
\hat U_{{\rm{OAT}}}^\dag {{\hat f}_ \pm }{{\hat U}_{{\rm{OAT}}}} = {{\rm{e}}^{ \pm i\chi t\left( {2{{\hat f}_z} \mp 1} \right)}}{{\hat f}_\pm }.\label{appeq1}
\end{align}
Using \(\hat f_y=(\hat f_+ - \hat f_-)/(2i)\), we obtain
\begin{eqnarray}
\hat U_{{\rm{OAT}}}^\dag {{\hat f}_y}{{\hat U}_{{\rm{OAT}}}} &=& \frac{1}{{2i}}\hat U_{{\rm{OAT}}}^\dag \left( {{{\hat f}_ + } - {{\hat f}_ - }} \right){{\hat U}_{{\rm{OAT}}}}\nonumber\\
 &=& \frac{1}{{2i}}\left[ {{{\rm{e}}^{i\chi t\left( {2{{\hat f}_z} - 1} \right)}}{{\hat f}_ + } - {{\rm{e}}^{ - i\chi t\left( {2{{\hat f}_z} + 1} \right)}}{{\hat f}_ - }} \right]\nonumber\\
 &=& \frac{1}{{2i}}\left[ {{{\rm{e}}^{i\chi t\left( {2{{\hat f}_z} - 1} \right)}}\left( {{{\hat f}_x} + i{{\hat f}_y}} \right)} \right.\nonumber\\
&&\left. { - {{\rm{e}}^{ - i\chi t\left( {2{{\hat f}_z} + 1} \right)}}\left( {{{\hat f}_x} - i{{\hat f}_y}} \right)} \right].\label{appeq2}
\end{eqnarray}
For the initial spin state \(\ket{f}\), one has
\begin{eqnarray}
\hat f_x \ket{f} = f \ket{f}.\label{appeq3}
\end{eqnarray}
Moreover, since \(\ket{f}=e^{-i\frac{\pi}{2}\hat f_y}\ket{f}^{z}\), where
\(\ket{f}^{z}\equiv \ket{f,m_f^z=f}\) is the eigenstate of \(\hat f_z\),
we obtain
\begin{align}
\hat f_y \ket{f}
&= \hat f_y e^{-i\frac{\pi}{2}\hat f_y}\ket{f}^{z}
= e^{-i\frac{\pi}{2}\hat f_y}\hat f_y \ket{f}^{z} \nonumber\\
&= e^{-i\frac{\pi}{2}\hat f_y}\frac{\hat f_+ - \hat f_-}{2i}\ket{f}^{z}\nonumber\\
&= \frac{i\sqrt{2f}}{2}\,e^{-i\frac{\pi}{2}\hat f_y}\ket{f-1}^{z}.
\end{align}
Noting that
\(\ket{f-1}= e^{-i\frac{\pi}{2}\hat f_y}\ket{f-1}^{z}\),
we finally have
\begin{align}
\hat f_y \ket{f} = i\sqrt{\frac{f}{2}}\,\ket{f-1}.\label{appeq5}
\end{align}
Using Eq. (\ref{appeq2}) together with the relations in Eqs. (\ref{appeq3}) and (\ref{appeq5}), we obtain the spin-down state as
\begin{align}
\hat U_{{\rm OAT}}^\dagger \hat f_y \hat U_{{\rm OAT}} \ket{f}
&= \frac{1}{2i}
   e^{i\chi t(2\hat f_z-1)}
   \Bigl(f\ket{f}-\sqrt{\frac{f}{2}}\ket{f-1}\Bigr) \nonumber\\
&\quad
   - \frac{1}{2i}
   e^{-i\chi t(2\hat f_z+1)}
   \Bigl(f\ket{f}+\sqrt{\frac{f}{2}}\ket{f-1}\Bigr) \nonumber\\
&= f e^{-i\chi t}\sin(2\chi t\hat f_z)\ket{f} \nonumber\\
   &+ i\sqrt{\frac{f}{2}}\,e^{-i\chi t}\cos(2\chi t\hat f_z)\ket{f-1},
\label{appeq6}
\end{align}
which is the result given in Eq. (\ref{replyeq7}) in the main text.

In the $f_z$ basis, the initial spin state can be written as
$\ket{f}=\sum_{m=-f}^{f}\mathcal{C}_m\ket{m}^z$,
where
$\mathcal{C}_m=\frac{1}{2^f}\binom{2f}{f+m}^{1/2}$ \cite{PhysRevA.47.5138}.
Accordingly, the state $\ket{f-1}$ can be expressed as
$\ket{f-1}=-\sqrt{\frac{2}{f}}\sum_{m=-f}^{f} m\,\mathcal{C}_m \ket{m}^z$. Then, the spin-down state is found to be
\begin{align}
\ket{\downarrow'}
&= e^{-i\chi t}\sum_{m=-f}^{f}\mathcal C_m
\left[
f\sin(2m\chi t)-i\,m\cos(2m\chi t)
\right]\ket{m}^z. \nonumber
\end{align}
Setting $\chi t=\pi/2$, we obtain
\begin{align}
\ket{\downarrow'}
&= e^{-i\frac{\pi}{2}}\sum_{m=-f}^{f}\mathcal C_m
\left[f\sin(m\pi)-i\,m\cos(m\pi)\right]\ket{m}^z.\label{appeq7}
\end{align}
For half-integer $f$, the magnetic quantum number $m$ takes only half-integer values. As a result,
\begin{align}
\cos(m\pi)&=0, \sin(m\pi)=ie^{-im\pi}.
\end{align}
Equation~(\ref{appeq7}) therefore reduces to
\begin{align}
\ket{\downarrow'}
&= f\sum_{m=-f}^{f}e^{-i\pi m}\mathcal C_m\ket{m}^z= fe^{-i\pi f}\ket{-f}.\label{appeq9}
\end{align}
Equation~(\ref{appeq9}) is the result presented in the main text.

\section{Generation of the GHZ state via OAT evolution\label{AppB}}

In this Appendix, we present the derivation of GHZ-state generation under OAT evolution.
Let us consider the OAT evolution
$\hat U_{\rm{OAT}_{\pi/2}}=
e^{-i\frac{\pi}{2}\hat f_z^2}$, corresponding to the Hamiltonian $\hat H_{\rm{OAT}}=\chi \hat f_z^2$ at time
$t=\pi/(2\chi)$.
For the initial CSS $\ket{f}$, we obtain the evolved state
\begin{align}
\hat U_{\rm{OAT}_{\pi/2}}\ket{f}
=
\sum_{m=-f}^{f} \mathcal{C}_m e^{-i\frac{\pi}{2}m^2}\ket{m}^z .\label{appeBq3}
\end{align}
For half-integer $f$, all magnetic quantum numbers $m$ are half-integers,
so one may write $m=n+1/2$ with $n\in\mathbb Z$. Then, we have
\begin{align}
m^2 = n(n+1)+\frac14 .
\end{align}
Since $n(n+1)$ is always even, one has
\begin{align}
e^{-i\frac{\pi}{2}m^2}
=
e^{-i\frac{\pi}{8}} e^{-i\frac{\pi}{2}n(n+1)}
=
e^{-i\frac{\pi}{8}}(-1)^{n(n+1)/2}.
\end{align}
On the other hand,
\begin{align}
e^{-i\frac{\pi m}{2}}+e^{i\frac{\pi m}{2}}
&=
2\cos\frac{\pi m}{2}\nonumber\\
&=
2\cos\left(\frac{\pi n}{2}+\frac{\pi}{4}\right)\nonumber\\
&=
\sqrt2\,(-1)^{n(n+1)/2},
\end{align}
and therefore
\begin{align}
e^{-i\frac{\pi}{2}m^2}
=
\frac{e^{-i\frac{\pi}{8}}}{\sqrt2}
\left(
e^{-i\frac{\pi m}{2}}+e^{i\frac{\pi m}{2}}
\right).\label{appeqB4}
\end{align}
Substituting Eq. (\ref{appeqB4}) into Eq. (\ref{appeBq3}) yields
\begin{align}
\hat U_{\rm{OAT}_{\pi/2}}\ket{f}
&=
\frac{e^{-i\frac{\pi}{8}}}{\sqrt2}
\sum_{m=-f}^{f} C_m
\left(
e^{-i\frac{\pi m}{2}}+e^{i\frac{\pi m}{2}}
\right)\ket{m}^z
\nonumber\\
&=
\frac{e^{-i\frac{\pi}{8}}}{\sqrt2}
\left(
e^{-i\frac{\pi}{2}\hat f_z}
+
e^{i\frac{\pi}{2}\hat f_z}
\right)\ket{f}.
\end{align}
The physical picture described by this expression is a superposition of two coherent spin states with opposite azimuthal angles--- i.e., rotated by $+\pi/2$ and $-\pi/2$ about the $z$-axis, respectively.
Next, we relate the rotated states to the $\hat f_y$ eigenstates by means of the following relations:
\begin{align}
e^{-i\frac{\pi}{2}\hat f_z}\ket{f}
=
e^{-i\pi \frac{f}{2}}\ket{f}^y,
e^{i\frac{\pi}{2}\hat f_z}\ket{f}
=
e^{i\pi \frac{f}{2}}\ket{-f}^y,\nonumber
\end{align}
where
\(\ket{f}^{y}\equiv \ket{f,m_f^y=f}\) is the eigenstate of \(\hat f_y\).
Hence,
\begin{align}
\hat U_{\rm{OAT}_{\pi/2}}\ket{f}
&=
\frac{e^{-i\frac{\pi}{8}}}{\sqrt2}
\left(
e^{-i\pi \frac{f}{2}}\ket{f}^y
+
e^{i\pi \frac{f}{2}}\ket{-f}^y
\right)
\nonumber\\
&=
\frac{e^{-i\pi(\frac{1}{8}+\frac{f}{2})}}{\sqrt2}
\left(
\ket{f}^y + e^{i\pi f}\ket{-f}^y
\right).
\end{align}
Up to an overall phase, this is a GHZ state in the $\hat f_y$ basis,
with a relative phase $e^{i\pi f}$.

\section{Achieving inverse OAT evolution via detuning control\label{Appc}}
To realize inverse OAT evolution, one needs to reverse the sign of $a_2(\Delta)/\Delta$. In this Appendix, we derive how this can be achieved by controlling the detuning $\Delta$. As a concrete example, we consider $^{87}\mathrm{Rb}$, for which the tensor-polarizability coefficient is given by \cite{PhysRevLett.133.173604}
\begin{equation}
a_2(\Delta)=\frac{\sqrt{2}}{40}
\left(
\frac{1}{1-\Delta_{13}/\Delta}
-\frac{5}{1-\Delta_{23}/\Delta}
+4
\right),
\label{eq:a2_def}
\end{equation}
where $\Delta_{13}=2\pi\times 423.60~\mathrm{MHz}$ and $\Delta_{23}=2\pi\times 266.65~\mathrm{MHz}$ denote the hyperfine splittings between the excited-state sublevels $\ket{F'=1}$ and $\ket{F'=3}$, and between $\ket{F'=2}$ and $\ket{F'=3}$ of the $5^2P_{3/2}$ manifold, respectively. The detuning $\Delta$ is defined relative to the transition $5^2S_{1/2},F=2 \rightarrow 5^2P_{3/2},F'=3$, with $\Delta>0$ ($\Delta<0$) corresponding to red (blue) detuning. Thus, we have
\begin{equation}
\frac{a_2(\Delta)}{\Delta}
=
\frac{\sqrt{2}}{40}
\left(
\frac{1}{\Delta-\Delta_{13}}
-\frac{5}{\Delta-\Delta_{23}}
+\frac{4}{\Delta}
\right).
\label{eq:a2_over_delta}
\end{equation}
We seek the condition under which ${a_2(\Delta_1)}/{\Delta_1}$ and ${a_2(-\Delta_2)}/(-{\Delta_2})$ have opposite signs while being equal in magnitude, which requires
\begin{equation}
\frac{a_2(\Delta_1)}{\Delta_1}
=
\frac{a_2(-\Delta_2)}{\Delta_2},  ~~~\Delta_{1},\Delta_{2}\geq 0,
\label{eq:condition_start}
\end{equation}
Substituting Eq.~(\ref{eq:a2_over_delta}) into Eq.~(\ref{eq:condition_start}) gives
\begin{equation}
\frac{1}{\Delta_1-\Delta_{13}}
-\frac{5}{\Delta_1-\Delta_{23}}
+\frac{4}{\Delta_1}
=
\frac{1}{\Delta_2+\Delta_{13}}
-\frac{5}{\Delta_2+\Delta_{23}}
+\frac{4}{\Delta_2}.
\label{eq:general_detuning_condition}
\end{equation}
Equation~(\ref{eq:general_detuning_condition}) provides the general relation between the detuning $\Delta_1$ and $\Delta_2$.
Consider the special case $\Delta_1=\Delta_2\equiv\Delta>0$, corresponding to red and blue detunings with equal magnitude, for which Eq.~(\ref{eq:general_detuning_condition}) reduces to
\begin{equation}
\frac{1}{\Delta-\Delta_{13}}
-\frac{5}{\Delta-\Delta_{23}}
=
\frac{1}{\Delta+\Delta_{13}}
-\frac{5}{\Delta+\Delta_{23}}.\label{eqC5}
\end{equation}
Solving Eq.~(\ref{eqC5}) yields
\begin{equation}
\Delta^2=
\frac{\Delta_{13}\Delta_{23}\left(5\Delta_{13}-\Delta_{23}\right)}
{5\Delta_{23}-\Delta_{13}}.
\label{eqC6}
\end{equation}
Substituting $\Delta_{13}$ and $\Delta_{23}$ into Eq. (\ref{eqC6}), we find
\begin{equation}
\Delta \approx 2\pi\times 479.5~\mathrm{MHz}.
\end{equation}
Therefore, when the red and blue detunings are chosen to have the same magnitude, $\Delta=\pm 2\pi\times 479.5~\mathrm{MHz}$,
the quantities $a_2(\Delta)/\Delta$ and $a_2(-\Delta)/(-\Delta)$ are equal in magnitude and opposite in sign.

\section{Analysis of squeezing in the presence of noise\label{Appc}}

In the presence of spontaneous scattering during a probe pulse of duration $T$,
the atomic oscillator quadrature obeys the Langevin equation \cite{hammerer2006quantum,PhysRevA.72.052313,PhysRevLett.85.5643}
\begin{equation}
\frac{d}{dt} \hat X_A(t)
=
-\frac{\beta}{2T}\hat X_A(t)
+
\sqrt{\frac{\beta}{T}}\,\hat f_{XA}(t),
\label{AppDeq1}
\end{equation}
where $\beta$ is the atomic decay rate, and
$\hat f_{XA}(t)$ is a vacuum noise operator satisfying
\begin{equation}
\langle \hat f_{XA}(t)\hat f_{XA}(t')\rangle
=
\frac{1}{2}\delta(t-t').
\label{eq:fXA_corr_app}
\end{equation}
The formal solution of Eq. (\ref{AppDeq1}) is
\begin{equation}
\hat X_A(t)
=
e^{-\frac{\beta t}{2T}}\hat X_A(0)
+
\sqrt{\frac{\beta}{T}}
\int_0^t ds\,
e^{-\frac{\beta (t-s)}{2T}}\hat f_{XA}(s).
\label{eq:XA_solution_app}
\end{equation}
The dissipation of the probe-light quadrature originates from both atomic absorption and reflections by the photon detectors, which is modeled as
\begin{equation}
\hat X_L^{\mathrm{out}'}
=
\sqrt{1-\epsilon}\left(\hat X_L^{\mathrm{in}}+\zeta\kappa\bar X_A\right)
+
\sqrt{\epsilon}\,\hat F_{LX},
\label{AppDeq4}
\end{equation}
where $\epsilon$ denotes the damping coefficient and ${{\hat F}_{LX}}$ is the optical vacuum operator satisfying $\langle{{\hat F}_{LX}}^2\rangle=1/2$.
Eq. (\ref{AppDeq4}) indicates that the optical readout depends on the time-averaged atomic quadrature
\begin{equation}
\bar X_A
\equiv
\frac{1}{T}\int_0^T dt\,\hat X_A(t).
\label{eq:Xbar_def_app}
\end{equation}
Substituting Eq.~\eqref{eq:XA_solution_app} into Eq.~\eqref{eq:Xbar_def_app}, one obtains
\begin{equation}
\bar X_A
=
A\,\hat X_A(0)+\hat N_A,
\label{eq:Xbar_decomp_app}
\end{equation}
where
\begin{eqnarray}
A
=
\frac{1}{T}\int_0^T dt\,e^{-\frac{\beta t}{2T}}
=
\frac{2}{\beta}\left(1-e^{-\frac{\beta}{2}}\right)
\label{eq:A_factor_app}
\end{eqnarray}
and
\begin{equation}
\hat N_A
=
\frac{1}{T}\int_0^T dt\,
\sqrt{\frac{\beta}{T}}
\int_0^t ds\,
e^{-\frac{\beta (t-s)}{2T}}\hat f_{XA}(s).
\label{eq:NA_def_app}
\end{equation}
Exchanging the order of integration in Eq.~\eqref{eq:NA_def_app} gives
\begin{equation}
\hat N_A
=
{\sqrt{\frac{\beta}{T}}}
\int_0^T ds\,
\frac{1-e^{-\frac{\beta (T-s)}{2T}}}{\beta/2}\,
\hat f_{XA}(s).
\label{eq:NA_compact_app}
\end{equation}
From Eqs. (\ref{eq:XA_solution_app}) and (\ref{AppDeq4}), the atomic quadrature after feedback control is obtained as
\begin{eqnarray}
{\hat X}_A^{F}&=&{{\hat X}_A}(T) + g\hat X_L^{{\rm{ou}}{{\rm{t}}^\prime }} \nonumber\\&=& {e^{ - \frac{\beta }{2}}}{{\hat X}_A}(0) + \sqrt {\frac{\beta }{T}} \int_0^T d s{\mkern 1mu} {e^{ - \frac{{\beta(T - s)}}{2T }}}{{\hat f}_{XA}}(s)\nonumber\\
 &&+ g\left[ {\sqrt {1 -\epsilon } \left( {\hat X_L^{{\rm{in}}} + \zeta \kappa  {{\bar X}_A}} \right) + \sqrt\epsilon  {\mkern 1mu} {{\hat F}_{LX}}} \right].\label{AppDeq10}
\end{eqnarray}
Calculating the variance of ${\hat X}_A^{F}$ yields
\begin{eqnarray}
2{\mathop{\rm var}} \left( {\hat X_A^F} \right) &\simeq& {e^{ - \beta }}{\left( {1 + g\sqrt {1 -\epsilon } \zeta \kappa } \right)^2}\nonumber\\
 &&+ {g^2} + \frac{\beta }{T}\int_0^T d s\left[ {{e^{ - \frac{\beta }{{2T}}\left( {T - s} \right)}}} \right.\nonumber\\
&&{\left. { + g\sqrt {1 -\epsilon} \zeta \kappa \frac{{1 - {e^{ - \frac{\beta }{{2T}}(T - s)}}}}{{\beta /2}}} \right]^2}.\label{AppDeq11}
\end{eqnarray}
We note that photon loss reduces the coupling constant by a factor of \(\sqrt{1-\epsilon}\). Defining $\tilde{\kappa}=\sqrt{1-\epsilon}\zeta \kappa$ Eq. (\ref{AppDeq11}) can be reexpressed as
\begin{eqnarray}
2{\rm{var}}\left( {\hat X_A^F} \right) = C\left( \beta  \right){g^2} + B\left( \beta  \right)g + 1,\label{AppDeq12}
\end{eqnarray}
where
\begin{align}
B(\beta) &= 2\tilde{\kappa} e^{-\beta} + \frac{4\tilde{\kappa}}{\beta} \left( 1 - e^{-\beta/2} \right)^2, \\
C(\beta) &= 1 + \tilde{\kappa}^2 e^{-\beta} \nonumber\\&+ \frac{4{\tilde{\kappa}}^2}{\beta^2} \left[ \beta - 4(1 - e^{-\beta/2}) + (1 - e^{-\beta}) \right].
\end{align}
By performing a Taylor expansion for $\beta \ll 1$ up to $\mathcal{O}(\beta^2)$, the coefficients can be simplified as
\begin{align}
B(\beta) &= 2\tilde{\kappa} - \tilde{\kappa}\beta + \frac{1}{2}\tilde{\kappa}\beta^2 + \mathcal{O}(\beta^3), \\
C(\beta) &= 1+\tilde{\kappa}^2 - \frac{2}{3}\tilde{\kappa}^2\beta + \frac{3}{8}\tilde{\kappa}^2\beta^2 + \mathcal{O}(\beta^3).
\end{align}
From Eq. (\ref{AppDeq12}), the minimum variance $2{\rm{var}}( {\hat X_A^F} )_{\rm{min}}= 1 - B^2/(4C)$ is found to be
\begin{equation}
2{\rm{var}}\left( {\hat X_A^F} \right)_{\rm{min}}\approx \frac{1}{1+\tilde{\kappa}^2} + \beta \frac{\tilde{\kappa}^2(3+\tilde{\kappa}^2)}{3(1+\tilde{\kappa}^2)^2}+ \mathcal{O}(\beta^2).
\end{equation}
In the limit of large \(\tilde{\kappa}\), this expression reduces to
\begin{equation}
\xi^2\equiv2{\rm{var}}\left( {\hat X_A^F} \right)_{\rm{min}}\approx \frac{1}{1+(1-\epsilon)\zeta^2\kappa^2} + \frac{\beta}{3}.
\end{equation}
This is the result given in Eq. (\ref{Feq14}) in the main text.

\end{document}